\documentclass{iopart}
\usepackage{graphicx}
\usepackage{comment}
\newcommand{\bfeps}{\mbox{\boldmath$\epsilon$}}
\begin{document}

\title{Spatial dispersion in Casimir forces: A brief review}

\author{R Esquivel-Sirvent,$^1$ C Villarreal$^1$, W L Moch\'an,$^2$, A
  M Contreras-Reyes$^3$, and V B Svetovoy$^4$}

\address{$^1$Instituto de F\'{\i}sica, Universidad Nacional Aut\'onoma
de M\'exico, Apartado Postal 20-364, 01000 Distrito Federal, M\'exico.}

\address{$^2$Centro de Ciencias F\'{\i}sicas, Universidad Nacional
Aut\'onoma de M\'exico, Apartado Postal 48-3, 62251 Cuernavaca,
Morelos, M\'exico.}
\address{$^3$Department of Physics and Astronomy, University of
  Sussex, Brighton, East Sussex BN1 9QH, United Kingdom}

\address{$^4$MESA+ Research Institute, University of Twente, P.O. Box 217,
7500 AE Enschede, The Netherlands}
\begin{abstract} 
We present the basic principles of non-local optics in connection with
the calculation of the Casimir force between half-spaces and thin
films. 
At currently accessible distances $L$, 
non-local corrections amount to about half a percent, but they
increase roughly as $1/L$ at smaller separations. Self consistent models lead
to corrections with the opposite sign as models with abrupt
surfaces.
\end{abstract}

\eads{\mailto{RES <raul@fisica.unam.mx>}, \mailto{CV
<carlos@fisica.unam.mx>}, \mailto{WLM <mochan@fis.unam.mx>},
  \mailto{AMC <amc40@sussex.ac.uk>}, \mailto{VBS <svetovoy@yandex.ru>}}


\section{Introduction}

The recent  measurements of Casimir forces \cite{lamoreaux,
mohideen, capasso,iannuzzi,decca} and their comparison with theory
have made it necessary to consider in detail the electromagnetic response
of the involved materials.  In this paper we concentrate our attention
on the spatial dispersion of the response which leads to the so
called non-local effects.  The problem of non-locality in connection
with Casimir forces was pointed out by Kats \cite{kats}, who made a
qualitative estimate of the effect and concluded that it was necessary
to specify the correct dependence of the dielectric function on both
the frequency and the wave-vector.  A more formal study was done by
Heinrichs \cite{heinrichs} and Buhl \cite{buhl} who studied the Van
der Waals interaction taking into account spatial dispersion using a
hydrodynamic model for the electronic dynamics, showing that at large
distances non-local effects 
were negligible. More recently, the study of non-local effects in
Casimir forces was revived
\cite{mochan03,oklahoma,esquivel04,esquivel05,mochan05} showing 
the need for an accurate theoretical description of the
system. It has also been shown that the correct understanding of
spatial dispersion is fundamental in order to solve recent
controversies regarding the behavior of the Casimir force at finite
temperatures \cite{sernelius05,svetovoy05}.
In this paper we present a brief review of non-local effects, their
incorporation in the calculation of optical properties, and their
importance in Casimir forces at zero temperature,  with
particular attention to the importance of a proper optical model of
the materials.

\section{Non-Local Media}

In a linear and causal time-independent system, the most general
relation between the electric displacement $\vec D$ and the electric
field $\vec E$ is
\begin{equation}\label{constitutive1}
\vec D(\vec r,t)=\int d^3r' \int_{-\infty}^{t} dt' \,
\bfeps(\vec r,\vec {r\,}',t-t') \cdot \vec E(\vec {r\,}',t'), 
\end{equation}
where $\bfeps(\vec r,\vec r',t-t')$ is the dielectric
response tensor of the material.
The response of the system at time $t$ depends on the excitation not
only at  $t$, but also at previous times $t'<t$, i.e., its
response is not instantaneous. This fact, known as temporal
dispersion, is closely related to the
well known dependence of the index of refraction on
frequency $\omega$. The time integration in Eq. (\ref{constitutive1})
is readily recognized as a convolution, so that it may be eliminated
through a temporal Fourier transform, 
\begin{equation}\label{constitutive2}
\vec D(\vec r,\omega)=\int d^3 r'\,  \bfeps(\vec r, \vec
{r\,}',\omega) \cdot \vec E(\vec {r\,}',\omega).
\end{equation}
Eq. (\ref{constitutive1}) also shows that the
response at a position $\vec r$ might depend not only on the excitation at
the same point, but also on positions $\vec {r\,}'$ within some
neighborhood $\Omega$ of $\vec r$. This non-local dependence arises from 
the interaction among different parts of the system and is known as
spatial dispersion due to its resemblance to
temporal dispersion. The size of the neighborhood $\Omega$ within
which $\bfeps(\vec r,\vec {r\,}',t-t')$ is non-negligible is
called the range of non-locality. 
Its size is typically about an atomic
distance, i.e., a few \AA, as that is the only length-scale that
characterizes the microscopic response of a material, although it may
become orders of magnitude larger, for example, in ultra-pure
conductors at low temperatures.
As the wavelength of light is typically much larger than
the range of non-locality, it is common to assume that $\vec E(\vec
{r\,}')\approx \vec E(\vec r)$ within 
$\Omega$, so that we may take the electric field out of the spatial
integration in Eqs. (\ref{constitutive1}) and (\ref{constitutive2}),
yielding a local response $\vec D(\vec r,\omega)=\bfeps(\vec
r,\omega) \cdot \vec E(\vec r,\omega)$, where $\bfeps(\vec
r,\omega)\equiv\int d^3r'\,\bfeps(\vec r, \vec
{r\,}',\omega)$ is the local dielectric response. An equivalent result
is obtained if we replace $\bfeps(\vec r,\vec {r\,}',\omega)$
by the local kernel $\bfeps(\vec r,\omega)\delta(\vec r-\vec
{r\,}')$ in Eq. (\ref{constitutive2}). However, we remark that this
approximation might fail close to the surface of a material where its
dielectric properties change rapidly, as some component of the field
have rapid variations in this region regardless of the
frequency. Thus, the detailed study of the electromagnetic screening at
surfaces requires a non-local approach. From the optical point of
view, non-locality produces corrections to the optical coefficients,
such as the reflection amplitudes, of order $\Lambda/\lambda$ where
$\Lambda$ is the characteristic length-scale of the selvedge region
where non-locality has to be accounted for \cite{halevi} and $\lambda$ is the
wavelength of light. As $\Lambda$ is typically a few
\AA, while $\lambda$ is on the order of hundreds or thousands of \AA,
non-locality may be safely ignored in many applications, although it
has to be accounted for in any precise calculation. 
Notice that within a non-local material we have 
to include the full tensorial character of the response even in the
isotropic case, as the separation $\vec r-\vec {r\,}'$ between
excitation and observation positions defines a particular direction in
space. 

Even within non-local media, Eq. (\ref{constitutive2}) may be further
simplified if the system is translationally invariant,
with a response that depends on the separation $\vec r-\vec {r\,}'$,
instead of being a function of both $\vec r$ and $\vec {r\,}'$.
In this case, Eq. (\ref{constitutive2}) is also a
convolution and may be rewritten as a simple algebraic relation,
\begin{equation}\label{DvsE}
\vec D(\vec k,\omega)=\bfeps(\vec k,\omega)\cdot \vec
E(\vec k,\omega),
\end{equation}
by taking spatial Fourier transforms with wave-vector $\vec k$.
Thus, spatial dispersion is frequently identified with a wave-vector
dependent dielectric response. Examples of spatially dispersive systems are
excitonic semi-conductors, where the spatial dispersion arises from
the momentum dependence of the excitonic energy. Another well known
non-local system is the 
conduction electron gas of a conductor. Here the non-locality arises
from the correlation hole that surrounds every electron, so that
excitation of one electron at some position affects the response of
the system within a neighborhood on the order of the Thomas-Fermi
screening distance. Furthermore, an electron
excited at one position may contribute to the response at a distance
of a mean free path away. 

The identification of spatial dispersion with a wave-vector dependent
response has led to some subtle but pervading confusion in the
literature when discussing the optical properties of non-local
systems,  as $\bfeps(\vec k,\omega)$ {\em is not even a well
defined quantity} close to a surface. As translational invariance is
necessarily lost, the response of the system has to be written fully
as in Eq. (\ref{constitutive2}), or be simplified at most to
\begin{equation}\label{constitutive3}
\vec D(z,\vec Q,\omega)=\int dz'\,  \bfeps(z,z',\vec
Q,\omega) \cdot \vec E(z',\vec Q,\omega),
\end{equation}
where we assumed that the surface is normal to the $z$ axis and we
took spatial Fourier transforms with wave-vector $\vec Q$ along the
$x-y$ plane along which we may assume a 2D translational
invariance. Nevertheless, specific meaning can sometimes be
given to $\bfeps(\vec k,\omega)$ close to a surface, but only
under some additional simplifying assumptions about the nature of the
response, some of which will be discussed below. These assumptions and
their approximate nature should not be ignored. 

\section{Homogeneous systems}

Within a homogeneous isotropic material, the only preferred direction
is determined by  $\vec k$, so that the dielectric tensor may be
written as
\begin{equation}.\label{eLT}
\fl\qquad
	\bfeps(\vec k,\omega)= \epsilon_l(\vec
k,\omega) \mathbf P_l(\vec k) + \epsilon_t(\vec
k,\omega) \mathbf P_t(\vec k) = \epsilon_l(\vec
k,\omega)\frac{\vec k \vec k}{k^2} + \epsilon_t(\vec
k,\omega)\left(\mathbf 1-\frac{\vec k\vec k}{k^2}\right),
\end{equation}
where $\mathbf P_l(\vec k)$ and $\mathbf P_t(\vec k)$ are the
longitudinal and the transverse projectors.
Thus, for longitudinal and for transverse
exciting fields, the response may be taken as scalar. However, the
longitudinal dielectric function $\epsilon_l(\vec k,\omega)$ is in general
different from the transverse dielectric function $\epsilon_t(\vec
k,\omega)$, although they coincide among themselves and with the local
dielectric function in the $k\to 0$ limit.  

Substituting Eq. (\ref{eLT}) into (\ref{DvsE}) and the resulting
displacement into Maxwell equations for non-magnetic systems we obtain
\begin{equation}\label{Maxwella}
\vec k\times \vec E(\vec k,\omega)= \frac{\omega}{c} \vec B,
\end{equation}
\begin{equation}\label{Maxwellb}
\vec k\times\vec B(\vec k,\omega) = -\frac{\omega}{c} \left (\epsilon_l(\vec
k,\omega) \vec E_l(\vec k,\omega) + \epsilon_t(\vec k,\omega) \vec
E_t(\vec k,\omega)\right),
\end{equation}
where $\vec E_\alpha=\mathbf P_\alpha\cdot \vec E$ ($\alpha=l,t$).
Taking as usual the vector product of Eq. (\ref{Maxwellb}) with $\vec
k$ and substituting (\ref{Maxwella})  we obtain the dispersion
relation for transverse waves ($E_t\ne 0 $)
\begin{equation}\label{kt}
k^2=\frac{\omega^2}{c^2}\epsilon_t(\vec k,\omega),
\end{equation}
which differs from the usual local result only by the explicit
dependence of $\epsilon_t$ on $\vec k$. On the other hand, taking the
longitudinal projection of Eq. (\ref{Maxwellb}) we obtain that there
might also be free longitudinal fields ($E_l\ne 0$) within the bulk,
provided the 
wave-vector and frequency satisfy the longitudinal dispersion relation,
given implicitly by 
\begin{equation}\label{kl}
\epsilon_l(\vec k,\omega)=0.
\end{equation}
Both Eqs. (\ref{kt}) and (\ref{kl}) can provide more solutions within
non-local media than the  usual two independent transverse modes that
may be sustained by local materials.

The specific form of the dielectric function depends on the nature
of the material, dielectric or metallic, and the model used. For
example, a simple
model for semiconductors close to an excitonic transition
is that of a Lorentz oscillator \cite{goyo},
\begin{equation}\label{excitonic}
\epsilon_t(\vec k, \omega)=\epsilon_l(\vec k, \omega)=\epsilon_\infty
+ \frac{\omega_p^2}{\omega_T^2(k)-\omega^2-i\gamma\omega},
\end{equation}
with weight $\omega_p^2$, dissipation constant $\gamma$ and with a
wave-vector dependent resonance  energy
$\hbar \omega_T(k)=E_g-E_b+K$ that incorporates
the energy required to create an electron-hole pair given by the
energy gap  $E_g$ of the semiconductor,  the binding energy $E_b$ of
the exciton and its kinetic energy $K=\hbar^2 k^2/2 M$, where $M$ is
the excitonic mass. Here $\epsilon_\infty$ is the contribution from
the other, non-resonant transitions. An analogous simple model for
metals is the hydrodynamic model, with a local transverse response
$\epsilon_t(\omega)=1-\omega_p^2/(\omega^2+i\gamma\omega)$ given by
the Drude model, and a non-local longitudinal response
\begin{equation}\label{elong} 
\epsilon_l(\vec k ,\omega)=1-\frac{\omega_p^2}{\omega^2+i \omega
\gamma-\beta^2 k^2}.
\end{equation} 
Here, the spatial dispersion arises from the fact that electrons are
Fermions and so Pauli's principle implies that it takes energy to
increase their density. Therefore, for longitudinal waves there is a
restoring force proportional to $\beta^2=3 v_F^2/5$ \cite{cottam} in
addition to the electrical coupling to the electromagnetic field
\cite{barton}. Here, $v_F$ is the Fermi velocity and $\beta$ is
related to the compressibility of the metal.

More elaborate expressions may be obtained through a purely quantum
mechanical approach using linear response theory. One of such
approaches is the random phase approximation (RPA), in which the response of
the electron gas to the self-consistent oscillating electric field is
identified with the response of a gas of independent Fermions to
an external perturbing field. The response may then be found from
Kubo's formulae through the density-density and the current-current
equilibrium correlation functions of the Fermion gas, resulting in the 
Lindhard longitudinal dielectric function \cite{lindhard,ziman,kittel}
\begin{equation}\label{lind}
\epsilon_{l}(\vec  k,\omega)=1+\frac{3 \omega_p^2}{k^2 v_F^2}f_l,
\end{equation}
where
\begin{equation}\label{Lind}
\fl
f_l=\frac{1}{2}+\frac{1}{8w}\left[ [1-(w-u)^2]\ln \left( \frac{w-u+1}{w-u-1}\right)+
[1-(w+u)^2 ln \left( \frac{w+u+1}{w+u-1}\right)\right],
\end{equation}
$w=k/2k_F$, $u=\omega/k v_F$, and $k_F$ is Fermi's
wave-vector. Similar expressions have been 
obtained for the non-local transverse dielectric function
\cite{esquivel04}, including additional dissipation channels \cite{ford}.

\section{Surface}

       It is important to emphasize that a dielectric response of the
form $\bfeps(\vec k,\omega)$ may only be defined within the bulk of a
translational invariant system. When a surface is present, this
invariance is broken along its normal direction and the full response
$\bfeps(z,z',\vec 
Q,\omega)$ has to be employed. This may be obtained from
microscopic surface screening calculations that include the
self-consistent confining potential \cite{ansgar}. Alternatively,
simplified models may be obtained by writing the dielectric
response close to the surface in terms of the bulk response,
introducing simplifying assumptions about the interaction of electrons
with the surface.  The latter calculations have the added difficulty
of having to account for the possible excitation of longitudinal and/or
additional transverse waves given by all the real and
complex solutions of Eqs. (\ref{kt}) and (\ref{kl}), beyond the usual
two transverse modes of local optics. 

As an example, consider $p$ polarized light incident on the surface of
a metal described by the hydrodynamic model (\ref{elong}). Within the
bulk there is one $p$-polarized  transverse and one longitudinal
transmitted mode. Assuming that
these modes may be extrapolated up to the surface, which is taken as a
sharp discontinuity in the dielectric properties, the optical problem
is reduced to the calculation of the amplitude of the reflected wave,
the transmitted transverse wave and the longitudinal wave. Thus, three
boundary conditions are required. Maxwell's equations provide only two
independent conditions, so that the problem seems to be
under-determined and additional boundary conditions (ABC's) of
non-electromagnetic origin are called for. The problem arises of
course from the assumption that the response is bulk-like up to the
surface. Nevertheless, it is reasonable to assume that at the surface
of a non-local metal {\em all of the components of all the fields
ought to be continuous} \cite{forstmann}, not only the usual field
components 
$E_\|$, $H_\|$, $D_\perp$ and $B_\perp$, where $\|$ and $\perp$ denote
parallel and perpendicular. For instance, a discontinuity in the normal
component of the electric field $E_\perp$ would imply an infinite
charge density at the surface, which would unrealistically require an
infinite amount of energy according to Pauli's principle. Using this
ABC, one can obtain the reflection amplitude
\begin{equation}\label{rhidro}
r_p=\frac{\epsilon_t k_v-k_t+Q^2(\epsilon_t-1)/k_l}{{\epsilon_t k_v+k_t-Q^2(\epsilon_t-1)/k_l}}, 
\end{equation}
which differs from the local result $ r_p=(\epsilon_t
k_v-k_t)/(\epsilon_t k_v+k_t)$ due to the excitation of longitudinal
waves at the surface. Here, $k_v$, $k_t$, and $k_l$ are the
normal components of the wave-vector of the incident wave in vacuum, of
the transverse wave in the metal and of the longitudinal wave
respectively, for given 
values of $\vec Q$ and  $\omega$.

The hydrodynamic result above may be obtained as a particular case the
semi-classical infinite barrier model (SCIB) 
\cite{kliewer}. In this model it is assumed that there are two ways in
which an electron excited at $\vec {r\,}'$ within a semi-infinite
conductor occupying the half-space $z>0$ may propagate and
contribute at the response at $\vec r$: either it travels directly
from $\vec {r\,}'$ to $\vec r$ or else, it first propagates to some
point on the surface $z=0$ where it is specularly reflected back
into $\vec r$. Assuming no electrons are inelastically
nor diffusively scattered  at the surface, and ignoring the quantum
interference between incident and reflected electronic wave-functions,
the polarization at $\vec r$ within the {\em real} semi-infinite conductor
would be indistinguishable from the polarization in a {\em fictitious}
infinite system, provided in the latter we impose the specular
$z\leftrightarrow -z$ symmetry. 
For each electron at $z>0$ moving with speed $-v_z$ towards the
surface, there would be another electron in the fictitious system at $z<0$
moving with speed 
$v_z>0$; when the former reaches $z=0$ and leaves the $z>0$ half-space,
the latter enters the half-space, as if the
former were specularly reflected. In the fictitious system, $\vec
E_\|$ must be an even function of $z$ while $E_\perp$ must be an odd
function. All other vectorial quantities must have the same behavior,
while pseudo-vectors such as $\vec B$ and $\vec H$ must have the
opposite parity. In particular, the fields $\vec H_\|$
and $D_\perp$ are odd functions of $z$ and are therefore discontinuous
at the surface.  This may seem surprising until we notice that {\em
this discontinuity applies only to the fields of the fictitious, not of the
real system}, and that fictitious and real fields coincide only in
the half space $z>0$. According to Maxwell equations, the fictitious
fields have a singular source given by a fictitious external surface current
flowing at $z=0$. By calculating the fields produced within an
homogeneous metal by such a singular current we may obtain the surface
impedances  \cite{stratton}
\begin{equation}\label{Zs}
Z_s\equiv -\frac{E_{y}\left( +0\right) }{%
H_{x}\left( +0\right) }=\frac{i}{\pi }\frac{\omega
}{c}\int\limits_{-\infty }^{\infty }\frac{dk_{z}}{\left( \omega
^{2}/c^{2}\right) \epsilon _{t}-k^{2}},
\end{equation}
and 
\begin{equation}\label{Zp}
Z_p\equiv
\frac{E_{x}\left( +0\right) }{%
H_{y}\left( +0\right) }=\frac{i}{\pi }\frac{\omega
}{c}\int\limits_{-\infty }^{\infty }\frac{dk_{z}}{k^{2}}\left[
\frac{k_{x}^{2}}{\left( \omega ^{2}/c^{2}\right) \epsilon
_{l}}+\frac{k_{z}^{2}}{\left( \omega ^{2}/c^{2}\right) \epsilon
_{t}-k^{2}}\right],
\end{equation}
which are independent of the unspecified magnitude and phase of the
external current. {\em As in the real system there are no 
singularities and the fields are continuous across the boundary}, we
can write $Z_s=-E_y(0^-)/H_x(0^-)$ and $Z_p=E_x(0^-)/H_y(0^-)$ in
terms of the fields $\vec E(0^-)$ and $\vec H(0^-)$ in vacuum, which
may be written in terms of the  incident and a reflected waves, so
that we may solve for the reflection amplitudes
\begin{equation}
\label{refl}
r_s =\frac{Z_s - Z_{vs}}{Z_s + Z_{vs}},\quad r_p =\frac{Z_{vp} -
Z_p}{Z_{vp} + Z_p}, 
\end{equation} 
where $Z_{vs}=\omega/(k_v c)$ and $Z_{vp}=(k_v c)/\omega$ are the
vacuum surface impedances.

It can be easily shown that Fresnel's reflection amplitudes may be
obtained by substituting in Eqs. (\ref{Zs})-(\ref{refl}) the local,
wave-vector independent dielectric function. The excitation of
collective modes may be accounted for by substituting the dielectric
function (\ref{elong}), yielding the hydrodynamic result
(\ref{rhidro}). A full quantum mechanical bulk response such as
Lindhard's formulae (\ref{lind},\ref{Lind}) may also be employed,
accounting therefore also for electron-hole pair creation and for Landau
damping. Nevertheless, the SCIB results above are still not exact, as
they do not account for the microscopic nature of the surface, the
shape of the confining surface potential, the quantum oscillations
of the equilibrium and the induced density close to the surface.

Besides applying simplified models that approximate the surface response in
terms of the bulk response of the system, it is possible to obtain the
microscopic response of the surface $\bfeps(z,z',\vec Q,\omega)$
through the use of linear response theory. The most simple microscopic
model for metallic surfaces is the jellium model, in which electrons
are added to a homogeneous semi-infinite positive background. Using
density functional theory in the local density approximation (LDA) the
self-consistent confining potential,  electronic wave-functions and
equilibrium density profile 
may be obtained through a solution of the Kohn-Sham equations
\cite{kohn}. Through a generalization of the RPA known as the time
dependent LDA (TDLDA), the surface susceptibility and dielectric
response may be 
obtained \cite{ansgar}. Eq. (\ref{constitutive3}) is an integral
relation between 
$\vec D$ and $\vec E$. As the width of the 
selvedge region is usually small \cite{halevi} with respect to the
relevant optical
wavelength, the integro-differential Maxwell's equations may be solved
using a long-wavelength approximation \cite{perturbative}. For example, the
reflection amplitude for $p$ polarized light may be simply expressed
as \cite{feibelman}
\begin{equation}\label{rpvsd}
r_p= r_p^0\left[1+\frac{2i k_v \epsilon_t}{1+\epsilon_t k_v^2/Q^2} d_\perp
\right],
\end{equation} 
where $r_p^0$ is the non-perturbed reflection amplitude given by
Fresnel's relations, and
\begin{equation}\label{dperp}
d_\perp \equiv \frac{\int dz\, z \delta \rho(z)}{\int dz\, \delta \rho(z)}
\end{equation}   
is the position of the centroid of the distribution of charge
$\delta\rho$ induced at the surface of a metal in order to screen the
normal component of the electric field. As an illustrative example,
$d_\perp=-i/k_l$ within the hydrodynamic model.  It is then simply shown
that Eq. (\ref{rpvsd}) is consistent with (\ref{rhidro}) in the
long-wavelength limit. Nevertheless, $d_\perp$ has been calculated
within the TDLDA for many metals and its low frequency value has been
tabulated \cite{ansgar}.

\section{Casimir forces between non-local media}   

In the previous sections we have discussed the calculation of the
electromagnetic response and the optical properties of non-local
systems. These may be immediately related to the calculation of the
Casimir force between spatially dispersive media by noticing that
Lifshitz formula, when written in terms of the reflection amplitudes,
\begin{equation}\label{F}
\fl 
F(L)={\cal A}\frac{\hbar c}{2 \pi ^2}  \mbox{Re} \int_0^{\infty} dQ Q
\int_{q\ge 0} dk_v  
\frac{\tilde k_v^2}{q} \left[
\frac{r^{(1)}_s r^{(2)}_s e^{2i\tilde k_vL}}{1-r^{(1)}_s r^{(2)}_s e^{2i\tilde k_v L}} 
+ \frac{r^{(1)}_p r^{(2)}_p e^{2i\tilde k_v L}}{1-r^{(1)}_p r^{(2)}_p
e^{2i\tilde k_v L}}  
\right],
\end{equation}
is applicable to a wide class of systems, including homogeneous,
inhomogeneous, semi-infinite or finite, insulator or metallic,
dissipationless or absorptive, and local or spatially dispersive media
\cite{mochan03,oklahoma,esquivel04,esquivel05,mochan05,reynaud}. 
Here, $\tilde k_v=k_v+i\eta$ where
$\eta\to0^+$ is a positive infinitesimal and 
the integral over $k_v$ runs from $iQ$ to 0 and then to $\infty$, so
that $q=\omega/c$ remains real and positive, although the integration
trajectory may be manipulated into a more convenient one over the
imaginary axis, and we assumed the zero temperature case. 
The reason for the generality of Eq. (\ref{F}) is that
$\alpha$-polarized photons ($\alpha=s,p$) that
are not reflected coherently at the $a$-th wall ($a=1,2$) of the cavity with
amplitude $r^{(a)}_\alpha$ are lost from the cavity with probability
$1-|r^{(a)}_\alpha|^2$. However, detailed balance in thermodynamic
equilibrium implies that those photons are replaced by similar photons
through incoherent radiation from the walls or by being 
transmitted from the vacuum region beyond the system, at the same rate
as they are lost. Thus, both the 
coherent and incoherent contributions to the field are determined
by the same reflection amplitudes. By using an ancillary system with the
same optical coefficients as the real cavity walls the generality of
Eq. (\ref{F}) has been proved \cite{mochan03} for a wide class of
isotropic systems,  and it has recently been generalized to
anisotropic \cite{torres} and to fermion mediated interactions
\cite{manuel}.

\begin{figure}
\begin{center}
\scalebox{1}{%
\input{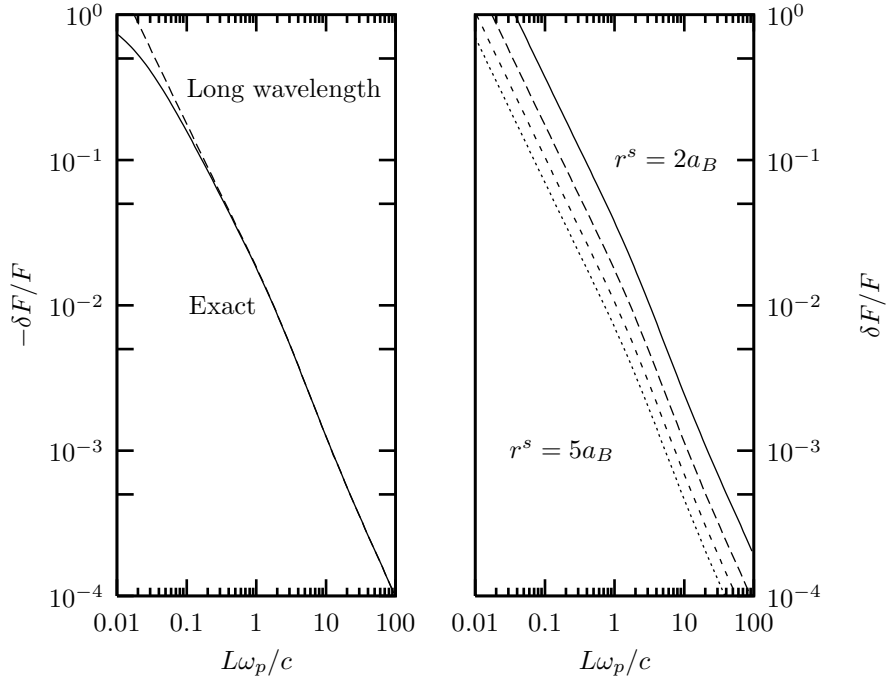}%
}%
\end{center}
\caption{\label{dF_F_allrs} Normalized non-local corrections $\delta
 F/F$ to the Casimir force as a
function of distance $L$ calculated for Au using the hydrodynamic model (left
panel) and for free electron metals of different densities corresponding
to $r^s/a_B=2$, 3, 4, and 5 using the self-consistent semi-infinite jellium
model (right panel), where $a_B$ is Bohr's radius and $r_s$ is related
to the electronic number density $n=3/(4\pi r_s^3)$.}
\end{figure}

Non-local effects in the Casimir force can therefore be obtained
quantitatively simply by substituting the appropriate non-local
reflection amplitudes in Eq. (\ref{F}). 
For example, in Fig. \ref{dF_F_allrs} we show the non-local corrections
$\delta F/F\equiv (|F_{nl}|-|F_l|)/|F_l|$
to the Casimir force calculated with the hydrodynamic and the self
consistent jellium models \cite{mochan05}, where the subscripts $nl$
and $l$ denote non-local and local respectively.
The exactly solvable hydrodynamic model predicts that non-locality
decreases the force due to the excitation of additional waves in the
media \cite{mochan03}. At the closest distances for which Casimir
forces have been measured, $L\sim 50$nm, the non-local correction
$|\delta F/F|$ is about half a percent. Similar results are also obtained
\cite{esquivel04} from models that employ more sophisticated
bulk dielectric response, such as a Lindhard-type dielectric function in
the region of anomalous dispersion, including a correction to account
for inter-band transitions, but which nevertheless truncate the system
abruptly at 
the surface employing the SCIB or similar models. As in the
hydrodynamic model, the force is smaller than in the local case.  
The left panel of Fig.  \ref{dF_F_allrs} also illustrates the accuracy
of the long wavelength approximation (Eqs. (\ref{rpvsd})). 
Contrariwise, the jellium model predicts 
a non-local correction of a similar size but of the opposite sign
(notice the labelling), that
is, non-local effects increase the magnitude of the Casimir force
\cite{mochan05}. The reason for this 
increase is that in realistic models of metallic surfaces, the
electronic density is not truncated abruptly at the position of the
nominal surface, but it decays smoothly to zero, extending beyond the
metal and into vacuum. Actually, the negative electronic charge
outside the nominal boundary and the compensating positive charge
within the metal form the surface dipole that is the source of the
self-consistent potential that actually confines the electrons within
the metal. It turns out that the region just outside of the metal is much more
polarizable than within the metal, so that most of the screening takes
place outside of the nominal surface. Thus, from the electromagnetic
point of view, the effective distance between two conductors is
smaller than the nominal distance, and therefore, the Casimir force is
larger.  Although the non-local correction $|\delta F/F|$ is relatively
small at currently accessible distances, it grows roughly as $L^{-1}$
and thus will becomes very important as smaller distances are
explored. Fig. \ref{dF_F_allrs} also shows that the non-local
correction increases with the electronic density. 

The role of thin metallic coatings in the calculation of Casimir
forces has also been studied taking into account spatial dispersion
\cite{esquivel05} within the Kliewer and Fuchs formalism. It was found
that the main non-local contributions come from the coupling of the
longitudinal guided collective modes of the thin films with
$p$-polarized light. Although it could have been expected that
non-local effects would be more important for thin films than for
semi-infinite media, as the width introduces an additional small
length-scale besides the relatively large wavelength, they were found
not to exceed about $7\%$ at small separations.
For current experimental setups and separations, non-local
corrections are on the order of $0.4\%$. The effect of thin films within
a local approximation has been explored by Lissanti \cite{lissanti05}.
   
\section{Conclusions}

We have reviewed the calculation of the optical properties of non-local
systems emphasizing some of the concepts that have frequently been a
source of confusion.  We have discussed the meaning of a
wave-vector dependent dielectric response when surfaces are present,
the problem of additional boundary conditions and the continuity
conditions for the fields. We have shown some expressions which may be
simply plugged into the Lifshitz formula in order to calculate the
Casimir force including non-local corrections. The effects of
non-locality are small at currently accessible distances but they
might become very important in future experiments that explore much
smaller distances. Self consistent theories produce a non-local
correction which has the opposite sign as that predicted by other
models, such as the hydrodynamic model, and more generally, the SCIB
model, in which the surface is unrealistically assumed to be truncated
abruptly. Thus, to obtain the correct sign, the microscopic electronic
density profile must not be disregarded. In this paper we have
concentrated in the non-local corrections at zero temperature and we
have not touched upon the important and controversial issue of the
thermodynamics of the Casimir force  at large separations,
where it is believed that non-locality also plays an important role
\cite{sernelius05}.

\ack This
work was partially supported  by DGAPA-UNAM under grants IN117402 and
No. IN118605, No. IN101605.  and CONACyT grant 44306.

\end{document}